\documentclass[a4,11pt]{cip-submit}  
\usepackage{afterpage}

\usepackage{mathptmx}
\usepackage{hyperref}
\hypersetup{
	colorlinks=false,
	pdfborder={0 0 0}
}
\def\Mr{\uppercase}

\usepackage{float}
\usepackage{wasysym}
\usepackage{graphicx,wrapfig,tikz}
\usepackage{caption}
\usepackage{subcaption}
\usepackage{amsmath,amsxtra,amssymb,latexsym,amscd,color,bm}
\usepackage[mathscr]{eucal}
\usepackage{epstopdf}
\usepackage{array}
\usepackage{cite}

\def\vsm{\vskip0.1cm}
\def\doi#1{\href{http://dx.doi.org/#1}{doi:#1}}

\def\titles#1{\title{\large\bf\noindent #1}}
\def\authors#1{\author{\begin{flushleft}{#1}\end{flushleft}}}
\def\authord#1#2{\indent\Mr{#1}\\
	\textit{\indent#2}\vsm}

\def\email#1{\bigskip\href{mailto:#1}{\textit{E-mail:}~{#1}}\\[3mm]}

\def\Keywords#1{\\[.2cm] Keywords:~{#1}.}

\def\and{$\text{\tiny AND }$}

\def\Classification#1{\\[.2cm] Classification numbers:~{#1}.} 
\begin{document}
	\Year{2019}
	\Page{1}\Endpage{15}
	\titles{No small hairs in anisotropic power-law Gauss-Bonnet  inflation}
	\authors{	\authord{Tuan Q. Do and Sonnet Hung Q. Nguyen}{Faculty of Physics, VNU University of Science,\\
			Vietnam National University, Hanoi 120000, Vietnam}
		\email{tuanqdo@vnu.edu.vn}
	}
	\maketitle
	\markboth{Tuan Q. Do and Sonnet Hung Q. Nguyen}{No small hairs in anisotropic power-law Gauss-Bonnet  inflation}

\begin{abstract}
We will examine whether anisotropic hairs exist in a string-inspired scalar-Gauss-Bonnet gravity model with the absence of potential of scalar field during the inflationary phase. As a result, we are able to obtain the Bianchi type I power-law solution to this model under the assumption that the scalar field acts as the phantom field, whose kinetic is negative definite. However,  the obtained anisotropic hair of this model turns out to be large, which is inconsistent with the observational data. We will therefore introduce a nontrivial coupling between scalar and vector fields such as $f^2(\phi)F_{\mu\nu}F^{\mu\nu}$ into the scalar-Gauss-Bonnet model with the expectation that the anisotropic hair would be reduced to a small one. Unfortunately, the magnitude of the obtained anisotropic hair is still large. These results indicate that the scalar-Gauss-Bonnet gravity model with the absence of potential of scalar field might not be suitable to generate  small anisotropic hairs during the inflationary phase. 
\Keywords{Gauss-Bonnet gravity, Inflation, Bianchi type I metric,  Cosmic no-hair conjecture}
\Classification{98.80.-k, 98.80.Cq, 98.80.Jk}
\end{abstract} 
\section{INTRODUCTION}\label{sec1}
 In cosmology, the  Copernican principle, which  states without any proofs that any spacetime describing the whole universe is just simply homogeneous and isotropic, has played a central role. According to this principle, the homogeneity and isotropy of universe should remain over its timeline. Testing the validity of the Copernican principle has been indeed a very important but not straightforward task for physicists and cosmologists  \cite{Saadeh:2016sak}.  Along with this principle, the has existed  the so-called cosmic no-hair conjecture proposed by Hawking and his colleagues also concerning the property of spacetime of universe \cite{GH}. In particular, this conjecture claims that a final state of our universe should be homogeneous and isotropic, regardless of any inhomogeneous and/or anisotropic initial states. This conjecture seems to be more general than the Copernican principle since it regards the evolution of universe from the past to the future. Unfortunately, a complete proof for this conjecture has been a great challenge to physicists and cosmologists for several decades. Of course, some partial proofs for this conjecture have been worked out \cite{wald,Barrow:1987ia,inhomogeneous,Carroll:2017kjo}.

Recently, the so-called cosmic inflation proposed by Guth and the others \cite{guth} to solve several classical puzzles such as the {\it flatness}, {\it horizon}, and {\it magnetic-monopole} problems, has emerged as one of leading paradigms  in the modern cosmology due to the fact many theoretical predictions of cosmic inflation are highly consistent with the observed data  of the Wilkinson Microwave Anisotropy Probe satellite (WMAP) ~\cite{WMAP} as well as the  Planck one~\cite{Planck}. Unfortunately,  some anomalies such as the hemispherical asymmetry and the cold spot of the cosmic microwave background (CMB) temperature, which have been firstly observed by the WMAP \cite{WMAP} and then confirmed by the Planck \cite{Planck}, cannot explained by standard inflationary models based on the Copernican principle. As a result, these exotic features imply that the state of the early universe might be anisotropic rather than isotropic. In cosmology, there exist  the so-called Bianchi spacetimes, which are known as homogeneous but anisotropic metrics and are divided  into nine types from type I to type IX ~\cite{Bianchi}. Hence, the Bianchi metrics could be useful in order to investigate the nature of the mentioned anomalies. It is worth noting that  some early works on the predictions of Bianchi inflationary era can be found in Ref. \cite{Pitrou:2008gk}, even when the anomalies was not detected.

It appears that the common thought that the early universe is just simply homogeneous and isotropic as described by  the Friedmann-Lemaitre-Robertson-Walker (FLRW) spacetime might no longer  be  valid \cite{Buchert:2015wwr}. Instead, we might think of a scenario that the  early universe might be described by the Bianchi spacetimes rather than the FLRW one since it might not be  isotropic but slightly anisotropic according to the data of the WMAP and Planck \cite{WMAP,Planck}. And it is noted that if the cosmic no-hair conjecture holds then the late time universe will be isotropic. However, the cosmic no-hair conjecture has faced a counter-example coming from a supergravity motivated model proposed by Kanno, Soda, and Watanabe (KSW) \cite{Soda,Soda1}, where  a unusual coupling of scalar and vector fields such as $f^2(\phi)F_{\mu\nu}F^{\mu\nu}$ is involved. As a result, the KSW model does admit  the Bianchi type I metric as its stable and attractor solution during the inflationary phase. More interestingly, this result still holds when a canonical scalar field $\phi$ is replaced by non-canonical ones, e.g., the Dirac-Born-Infeld, supersymmetric Dirac-Born-Infeld, and covariant Galileon scalar fields \cite{WFK}. Hence, the cosmic no-hair conjecture seems to be  violated extensively in the context of the KSW model.  As a result, the most important part of the KSW model is the coupling $f^2(\phi)F_{\mu\nu}F^{\mu\nu}$. It does play a leading role in  breaking down the validity of the cosmic no-hair conjecture. Consequently, there have been a number of papers investigating possible extensions of the KSW model to seek more counter-examples to the cosmic no-hair conjecture  \cite{WFK1,extensions}. It is therefore important to test the validity of the cosmic no-hair conjectures in the existing cosmological models.

It is noted that all the mentioned models have not discussed the effect of higher curvature terms such as the Gauss-Bonnet term \cite{primary}. It would be very interesting if such higher curvature terms could either support or break down the validity of the cosmic no-hair conjecture. Note that the Gauss-Bonnet term has been discussed extensively in the literature \cite{Lovelock:1971yv,ART,KSS,bianchi-I,darkenergy,review-darkenergy,inflation,observational-inflation,Lahiri,Antoniou:2017acq,blackhole,higherdim-blackhole,wormhole,total-derivative,no-go}. For example, it might provide us alternative approachs to some important cosmological problems such as the dark energy \cite{darkenergy,review-darkenergy} or the cosmic inflation \cite{inflation,observational-inflation}. In addition, some non-trivial solutions such as black holes \cite{Antoniou:2017acq,blackhole,higherdim-blackhole} and wormholes \cite{wormhole} have been shown to exist in the context of the Gauss-Bonnet gravity. More interestingly, anisotropic inflation has been claimed to exist in the context of the scalar-vector-Gauss-Bonnet model, in which the coupling $f^2(\phi)F_{\mu\nu}F^{\mu\nu}$ and the potential of scalar field are both involved \cite{Lahiri}.  In many inflation models, the potential of scalar field $\phi$, i.e., $V(\phi)$, plays an important role in order to cause an inflationary phase.  In the scalar-Gauss-Bonnet model, however, isotropic inflationary solutions can be shown to exist even when $V(\phi)$ is ignored as claimed in Ref. \cite{inflation}. Therefore, it would be interesting to examine whether anisotropic inflationary solutions with small hairs exist in the scalar-Gauss-Bonnet gravity model with the absence of $V(\phi)$. Hence, this is the topic of study presented in this paper.

As a result, the present paper is organized as follows: (i) A brief introduction of this research has been given in Sec. \ref{sec1}. (ii) A scalar-Gauss-Bonnet model and its Bianchi type I anisotropic solution will be shown in Sec. \ref{sec2}. (iii)  Then,  an extended scenario of the scalar-Gauss-Bonnet model, in which  a nontrivial coupling between scalar and vector fields, $f^2(\phi)F_{\mu\nu}F^{\mu\nu}$, is involved, will be discussed in Sec. \ref{sec3} to see whether  small hairs appear. (iv) Finally, concluding remarks will be given in Sec. \ref{sec4}.
\section{SCALAR-GAUSS-BONNET MODEL} \label{sec2}
\subsection{The setup}
As a result, an action of a string-inspired Gauss-Bonnet term coupled to a scalar field $\phi$ model is given by ~\cite{inflation,Antoniou:2017acq}
\begin{equation}\label{c1.action}
S= \int{d^4 x \sqrt{-g}} \left[\frac{M_p^2}{2}R - \frac{\omega}{2}\partial_\mu \phi \partial^\mu \phi -\frac{h(\phi)}{8}G \right],
\end{equation}
where $M_p$ is the reduced Planck mass, while $\omega =+ 1$ or $-1$ for a canonical or phantom scalar field \cite{WFK1}, respectively. Here, the potential of scalar field $V(\phi)$ has been neglected in a sense that the last term in the above action can play as an effective potential of scalar field \cite{inflation,Antoniou:2017acq}. Of course, the other scenario of the Gauss-Bonnet inflation, in which $V(\phi)$ shows up, has also been discussed extensively, e.g., see Refs. \cite{observational-inflation,Lahiri}.

In the action (\ref{c1.action}), $G$ coupled to a function of scalar field $h(\phi)$ acts as the Gauss-Bonnet  invariant
term, whose definition is given by \cite{primary,Lovelock:1971yv,ART,KSS,bianchi-I,darkenergy,review-darkenergy,inflation,observational-inflation,Lahiri,Antoniou:2017acq,blackhole,higherdim-blackhole,wormhole,total-derivative}
\begin{equation}
G =R^2 - 4R_{\mu\nu}R^{\mu\nu}+R_{\mu\nu\rho\sigma} R^{\mu\nu\rho\sigma} .
\end{equation}
Note also that the sign in front of the coupling $h(\phi)G/8$ could be either positive or negative definite, depending on the studied models ~\cite{primary,Lovelock:1971yv,ART,KSS,bianchi-I,darkenergy,review-darkenergy,inflation,observational-inflation,Lahiri,Antoniou:2017acq,blackhole,higherdim-blackhole,wormhole,total-derivative}. The existence of $h(\phi)$ is necessary in order to ensure that the Gauss-Bonnet term $G$ will not disappear in  four-dimensional spacetimes. This is based on the fact that   the Gauss-Bonnet term $\sqrt{-g}G$  can be shown to be a total derivative in four dimensions, e.g., see Refs. \cite{darkenergy,total-derivative}. For details of this claim, one can read, e.g., Refs. \cite{inflation,total-derivative}. 

As a result, varying the action (\ref{c1.action}) with respect to the inverse metric $g^{\mu\nu}$ leads to the modified Einstein field equation \cite{darkenergy,inflation,Lahiri} (for additional details of the derivation, one can see interesting papers in Ref. \cite{total-derivative})
\begin{align}\label{c1.Einstein}
&M_p^2 \left(R_{\mu\nu}-\frac{1}{2}g_{\mu\nu}R \right) - \left(R_{\mu\sigma \nu \rho}-g_{\mu\nu}R_{\sigma\rho}\right) \nabla^{\sigma} \nabla^{\rho}h + \left(R_{\mu\nu}-\frac{1}{2}g_{\mu\nu}R\right) \square h \nonumber\\
& -R_{\sigma\nu} \nabla_\mu \nabla^\sigma  h -R_{\mu\rho}\nabla^\rho  \nabla_\nu h + \frac{1}{2}R \nabla_\mu\nabla_\nu h  - \omega \partial_\mu \phi \partial_\nu \phi +\frac{\omega}{2}g_{\mu\nu}\partial_\sigma \phi \partial^\sigma \phi   =0, 
\end{align}
where $\square \equiv \nabla_\mu \nabla^\mu$ is the d'Alembert operator and $\nabla_\mu$ is the covariant derivative.
 In addition, the corresponding field equation of the scalar field $\phi$ turns out to be
\begin{equation}\label{c1.scalar1}
\omega \square \phi =  \frac{\partial_\phi h}{8}G  ,
\end{equation}
where $\partial_\phi \equiv \partial/\partial\phi$ and $\square \equiv  \frac{1}{\sqrt{-g}} \partial_\mu (\sqrt{-g} \partial^{\mu})$.
For the full modified Einstein  equations of Gauss-Bonnet gravity, which contain more terms coupled to  $h(\phi)$ and hold in arbitrary dimensions, see Ref.  ~\cite{darkenergy}. It is straightforward to see that if we set $h(\phi)=\text{constant}$ then all Gauss-Bonnet terms in the Einstein field equations (\ref{c1.Einstein}) will vanish automatically. Hence, $h(\phi)$ should be non-constant, i.e., a function of scalar field in order to maintain the string effect  in the field equations in terms of the Gauss-Bonnet terms. For the constant-like $h(\phi)$ case, the only way to see the Gauss-Bonnet effect is working in high dimensions, e.g., five dimensions, where the Gauss-Bonnet terms no longer vanish~\cite{higherdim-blackhole}. 

In order to seek anisotropic hairs, we will work on the Bianchi type I metric ~\cite{Soda,WFK},
\begin{equation}\label{bianchi-I}
 ds^2 =  -dt^2+\exp\left[{2\alpha(t)-4\sigma(t)}\right]dx^2+\exp\left[{2\alpha(t)+2\sigma(t)}\right]\left({dy^2+dz^2}\right)  , 
\end{equation}
where $\alpha(t)$ acts as an isotropic parameter, while $\sigma(t)$ stands for a deviation from an isotropic space, which should be  much smaller than $\alpha(t)$ during the inflationary phase, i.e., $|\sigma(t)| \ll \alpha(t)$,  in order to be consistent with recent observations such as the WMAP ~\cite{WMAP} or Planck ~\cite{Planck}. Note that $\sigma(t)$ is not necessarily positive definite. However, $\alpha(t)$ must be positive definite since it plays as a leading role in the expansion rates of the universe. In other words, $\alpha-2\sigma >0$ and $\alpha +
\sigma>0$ are necessary constraints for expanding solutions. Note also that the Bianchi type I metric has also been studied in order to reveal the nature of singularities in the context of Gauss-Bonnet model ~\cite{bianchi-I}. 
 
As a result, the corresponding non-vanishing components of Einstein field equation (\ref{c1.Einstein}) can be defined to be  (see the Appendix  for the derivations)
\begin{align}
\label{c1.alpha1}
\dot\alpha^2 &= \dot\sigma^2 +\frac{\omega \dot\phi^2}{6M_p^2} +\frac{\dot h}{M_p^2} \left(\dot\alpha^3 -2\dot\sigma^3-3\dot\alpha\dot\sigma^2\right) , \\
\label{c1.alpha2}
\ddot\alpha &= -3\dot\alpha^2 +\frac{\dot h}{2M_p^2} \left(2\ddot\alpha \dot\alpha -2\ddot\sigma \dot\sigma +5\dot\alpha^3 -9\dot\alpha \dot\sigma^2 -4\dot\sigma^3\right)+\frac{\ddot h}{2M_p^2} \left(\dot\alpha^2-\dot\sigma^2\right) , \\
\label{c1.sigma}
\ddot\sigma &= -3\dot\alpha \dot\sigma +\frac{ \dot h}{M_p^2} \left[ \ddot\alpha \dot\sigma +\ddot\sigma \left(\dot\alpha+2\dot\sigma\right) + 3 \dot\alpha \dot\sigma \left(\dot\alpha+\dot\sigma\right)  \right] +\frac{\dot\sigma \ddot h }{M_p^2}\left(\dot\alpha+\dot\sigma  \right) .
\end{align}
These equations are consistent with that derived in Ref. \cite{Lahiri}.
It turns out that the last equation (\ref{c1.sigma}) characterizes the  evolution of anisotropy parameter $\sigma$. It is straightforward to see that if  $h(\phi)$ is set to be zero or constant, then a trivial solution of Eq.  (\ref{c1.sigma}) turns out to be $\sigma=0$, which corresponds to an isotropic universe. On the other hand, the following scalar field equation (\ref{c1.scalar1}) reads
\begin{equation} \label{c1.phi}
\omega \ddot\phi  = -3\omega \dot\alpha \dot\phi -3\left(\dot\alpha + \dot\sigma\right) \left[\ddot\alpha \left(\dot\alpha -\dot\sigma\right) -2\ddot\sigma \dot\sigma +\dot\alpha^3 -\dot\alpha\dot\sigma \left(\dot\alpha+2\dot\sigma\right)\right]\partial_\phi h  ,
\end{equation}
where we have used the explicit definition of $G$ shown in Eq. (\ref{RGB}) in the Appendix,
\begin{equation}
G = 24 \left(\dot\alpha + \dot\sigma\right) \left[\ddot\alpha \left(\dot\alpha -\dot\sigma\right) -2\ddot\sigma \dot\sigma +\dot\alpha^3 -\dot\alpha\dot\sigma \left(\dot\alpha+2\dot\sigma\right)\right].
\end{equation}
\subsection{Anisotropic power-law solutions}
We will seek anisotropic power-law solutions with the following forms~\cite{Soda,WFK},
\begin{equation} \label{ansatz}
\alpha = \zeta \log \left( t \right),~ \sigma  = \eta \log \left( t \right),~ \frac{\phi }
{{M_p }}  = \xi \log \left( t \right) + \phi _0,
\end{equation}
along with the exponential function,
\begin{equation}
h(\phi) = h_{0} \exp\left[\frac{\lambda \phi}{M_p}\right].
\end{equation}
Here $\lambda$, $\phi_0$,  and $h_0$ are constants. In addition, $\lambda$  will be regarded as a positive parameter, while the sign of $h_{0}$ could be positive or negative definite depending on specific scenarios.  Note  that the choice of exponential function $h(\phi)$ has been made in many previous papers, e.g., see Refs. \cite{bianchi-I,observational-inflation,darkenergy,review-darkenergy,blackhole,wormhole}, while other types of $h(\phi)$ such as the power-law type can  be found in Refs. \cite{inflation,observational-inflation}. Additionally, the isotropic power-law inflation has been investigated in single field models, where the potential of scalar field $V(\phi)$ is introduced ~\cite{observational-inflation}.

As a result, the field equations (\ref{c1.alpha1}), (\ref{c1.alpha2}), (\ref{c1.sigma}), and (\ref{c1.phi}) can be reduced to the following algebraic equations,
\begin{align}\label{c1.eq-Friedmann}
\zeta^2 &=  \eta^2 +\frac{\omega \xi^2}{6} +2 \left(\zeta^3 -2\eta^3 -3\zeta \eta^2\right)  u, \\
\label{c1.eq-zeta}
-\zeta &= -3\zeta^2 +  \left(5\zeta^3 -4\eta^3- 9\zeta\eta^2-2\zeta^2+2\eta^2 \right) u  + \left(\zeta^2 -\eta^2\right)u , \\
\label{c1.eq-eta}
-\eta &=-3\zeta \eta + 2  \eta \left(\zeta+\eta\right) \left(3\zeta-1\right)u ,\\
\label{c1.eq-phi}
-\omega\xi &= -3\omega  \xi\zeta -3\lambda  \left(\zeta +\eta \right) \left[\zeta^3 -\zeta \left(\zeta-\eta \right) -\zeta\eta \left(\zeta+2\eta\right)+2\eta^2\right] u ,
\end{align}
where  $u$ an  additional variable is defined as
\begin{equation}
u = \frac{h_{0}}{M_p^2}\exp[\lambda \phi_0],
\end{equation}
along with the following constraint,
\begin{equation}\label{c.1constraint-1}
\lambda \xi =2, 
\end{equation}
which leads all field equations to have the same power in time, i.e., $t^{-2}$. It appears that we now end up with four algebraic equations for three independent variables, $\zeta$, $\eta$, and $u$. However, only three of field equations,  Eqs. (\ref{c1.eq-zeta}), (\ref{c1.eq-eta}), and (\ref{c1.eq-phi}),  turn out to be independent equations. The reason is that the first equation (\ref{c1.eq-Friedmann}) coming from the Friedmann equation (\ref{c1.alpha1}) acts as a constraint equation such that all found solutions of the rest equations must satisfy it consistently. 

As a result, solving Eq. (\ref{c1.eq-eta})  leads to 
\begin{equation}\label{c1.sol-u1}
u =\frac{1}{2 (\zeta +\eta )}.
\end{equation}
{\color{black}Furthermore, Eq. (\ref{c1.eq-zeta}) can be reduced to an equation of $\eta$,
\begin{equation}
(\zeta+\eta)(\zeta+4\eta-1)=0, 
\end{equation}
with the help of solution shown in Eq. (\ref{c1.sol-u1}). It is apparent that $\eta$ can be solved to be
\begin{equation}\label{c1.solution-eta}
\eta = \frac{1-\zeta}{4},
\end{equation}
here we have ignored the solution $\eta=-\zeta$ due to the requirement for the existence of $u$ defined in Eq. (\ref{c1.sol-u1}).
Plugging these solutions into Eq. (\ref{c1.eq-Friedmann}) as well as Eq. (\ref{c1.eq-phi}) leads to the corresponding equations of $\zeta$,
\begin{align}
9\zeta^2 -6\zeta -3+\frac{32\omega}{\lambda^2} & =0,\\
(3 \zeta -1)  \left[9\zeta^2 -6\zeta-3 +\frac{32\omega}{\lambda^2}\right]&=0,
\end{align}
respectively.}  As a result, non-trivial solutions of $\zeta$ can be solved to be
\begin{equation} \label{c1.solution-zeta}
\zeta_\pm = \frac{1}{3} \pm \frac{2}{3\lambda} \sqrt{\lambda^2 -8 \omega}. 
\end{equation}
Here, we have ignored the trivial solution $\zeta =1/3$ since we would like to seek inflationary solutions with $\zeta \gg 1$.  It is clear that $\zeta_\pm <1$ for all values of $\lambda$ if  the scalar field $\phi$ is chosen to be canonical, i.e., $\omega =+1$. They can only be used for expanding solutions rather than inflationary ones with $\zeta \gg 1$.
On the other hand, we can obtain inflationary solutions for the phantom field with $\omega =-1$ since the square root in Eq. (\ref{c1.solution-zeta}) can be arbitrarily larger than one assuming $\lambda \ll 1$. Indeed, it is straightforward to show that
\begin{equation}\label{c11.solution-zeta}
\zeta=\zeta_{+} =\frac{1}{3} + \frac{2}{3\lambda} \sqrt{\lambda^2 +8} \simeq \frac{4\sqrt{2}}{3\lambda} \gg 1
\end{equation}
 for $0<\lambda \ll 1$.  Note that a quite similar scenario, in which the phantom field is coupled to the Gauss-Bonnet model, has been discussed in Ref. \cite{observational-inflation}. 

{\color{black}Hence, given the solution $\zeta=\zeta_{+}$ with $\omega=-1$, we are able to determine the corresponding value of $\eta$ to be
\begin{equation} \label{c1.solution-eta-2}
\eta =\frac{1}{6}- \frac{1}{6\lambda}\sqrt{\lambda^2+8}.
\end{equation}}
{\color{black}Given the ansatz shown in Eq. (\ref{ansatz}), the scale factors of the Bianchi type I metric given in Eq. (\ref{bianchi-I}) turn out to be power-law functions of time, i.e.,
\begin{align}
\exp[{\alpha(t) +\sigma(t)}] =t^{\zeta +\eta};~ \exp[{\alpha(t) -2\sigma(t)}] =t^{\zeta -2\eta}.
\end{align}
It appears that for expanding universes, we just need $\zeta+\eta >0$ and $\zeta-2\eta >0$. For inflationary universes having a very fast expansion, however,  $\zeta$ and $\eta$ must obey the following constraints \cite{Soda}},
\begin{align}\label{c1.constraint-3}
\zeta+\eta \gg 1;~\zeta-2\eta \gg 1.
\end{align}
Note that these constraints do not mean that $\eta$ must be positive as $\zeta$. {\color{black} It turns out that these inflationary constraints can be easily fulfilled if $\lambda\ll 1$. Indeed, it is straightforward to see that 
\begin{equation}
\eta \simeq -\frac{\sqrt{2}}{3\lambda},
\end{equation}
provided that $\lambda \ll 1$. Consequently, we have the following results,
\begin{align}
 \zeta +\eta \simeq \frac{\sqrt{2}}{\lambda} \gg 1;~ \zeta -2\eta \simeq \frac{2\sqrt{2}}{\lambda} \gg 1,
\end{align}
which represent the inflationary solution as expected.
Unfortunately, the magnitude of the corresponding anisotropy parameter  turns out to be very large
\begin{equation} \label{ratio}
\left|\frac{\eta}{\zeta}\right| \simeq \frac{1}{4}=0.25.
\end{equation}
To be more specific, we will numerically plot below $\zeta+\eta$, $\zeta-2\eta$, and $|\eta/\zeta|$ as functions of the field parameter $\lambda$ using the exact solutions shown in Eqs. (\ref{c11.solution-zeta}) and (\ref{c1.solution-eta-2}).}
\begin{figure}[hbtp]
\begin{center}
{\includegraphics[height=45mm]{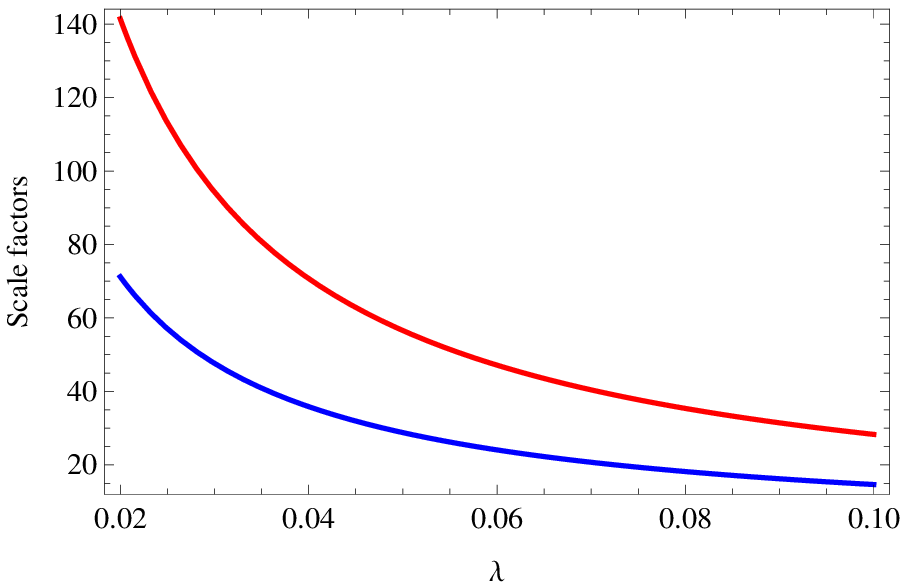}}\quad
{\includegraphics[height=45mm]{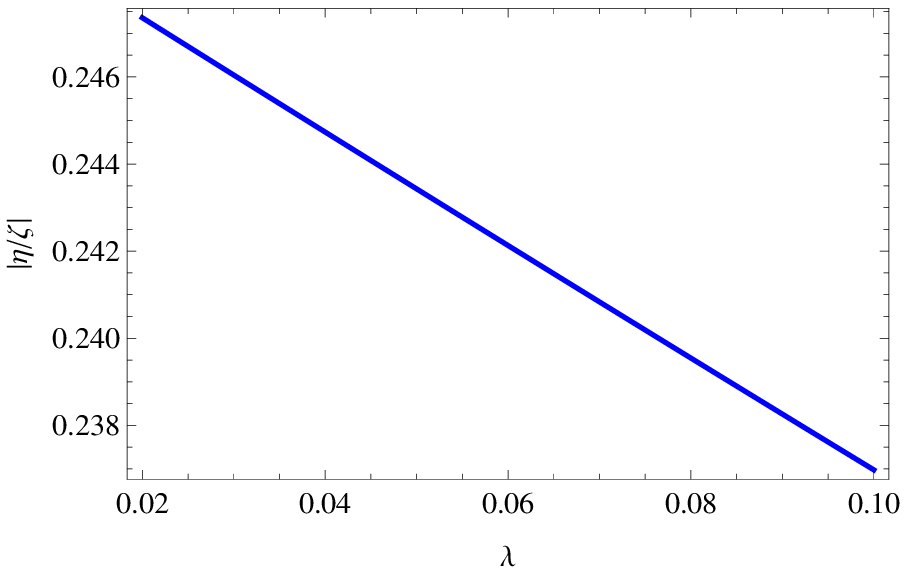}}\\
\caption{(Left) $\zeta-2\eta$ (upper red curve) and $\zeta+\eta$ (lower blue curve) as  functions of $\lambda$. (Right) $\left|\eta/\zeta\right|$ as a function of $\lambda$.} 
\label{fig3} 
\end{center}
\end{figure}
According to these plots, it appears that the smaller value of field parameter $\lambda$ is, the larger values of scale factors are and of course the larger anisotropy is. Additionally, the plots show that the obtained solution is highly anisotropic, while the anisotropic deviation parameter $\eta$ (or $\sigma$) is expected to be much smaller than the isotropic one $\zeta$ (or $\alpha$) in order to be consistent with the observational data of WMAP and Planck. This result implies that the Gauss-Bonnet term tends to significantly enhance the magnitude of anisotropic hairs rather than reduce them. Note that the similar effect has also been discovered in Ref. \cite{Lahiri}. Now, we would like to have anisotropic inflationary solutions having small hairs for the Gauss-Bonnet model. This problem might be solved by introducing extra field(s), e.g., the electromagnetic field into the action (\ref{c1.action}), according to the recent investigation in \cite{Lahiri}. Indeed, it   has pointed out in Ref. \cite{Lahiri} that anisotropic power-law inflations with a small anisotropy could be obtained within an extended framework of the Gauss-Bonnet model, in which a nontrivial coupling between scalar  and electromagnetic fields such as $f^2(\phi)F_{\mu\nu}F^{\mu\nu}$ \cite{Soda,WFK} is introduced. However, this model has been proposed to deal with a canonical scalar field along with its pure potential $V(\phi)$. Hence, it is interesting to see whether a small anisotropy still appears in a quite different scenario, in which $V(\phi)$ is absent \cite{inflation}. {\color{black}It is worth noting that the magnitude of the ratio $\eta/\zeta$ in models involving the coupling $f^2(\phi)F_{\mu\nu}F^{\mu\nu}$ would be ${\cal O}(10^{-9})$ in order to agree with the observational data of the Planck, according to a careful investigation in Ref. \cite{Naruko:2014bxa}}.  
\section{SCALAR-VECTOR-GAUSS-BONNET MODEL} \label{sec3}
As said above, we will consider the following action without the contribution of the pure potential of scalar field $V(\phi)$,
\begin{equation}\label{c2.action}
S= \int{d^4 x \sqrt{-g}} \left[\frac{M_p^2}{2}R - \frac{\omega}{2}\partial_\mu \phi \partial^\mu \phi -\frac{h(\phi)}{8}G -\frac{f^2(\phi)}{4}F_{\mu\nu}F^{\mu\nu} \right],
\end{equation}
where $f(\phi)$ is a function of $\phi$ and $F_{\mu\nu}  \equiv \partial_\mu A_\nu -\partial_\nu A_\mu$ is the field strength of the vector (electromagnetic) field $A_\mu$. Note that a different scenario, in which $V(\phi)$ is involved, has been investigated in Ref. \cite{Lahiri}. As a result, the corresponding field equations of this model turn out to be
\begin{align}\label{c2.Einstein}
&M_p^2 \left(R_{\mu\nu}-\frac{1}{2}g_{\mu\nu}R \right) - \left(R_{\mu\sigma \nu \rho}-g_{\mu\nu}R_{\sigma\rho}\right) \nabla^{\sigma} \nabla^{\rho}h + \left(R_{\mu\nu}-\frac{1}{2}g_{\mu\nu}R\right) \square h \nonumber\\
& -R_{\sigma\nu} \nabla_\mu \nabla^\sigma  h -R_{\mu\rho}\nabla^\rho  \nabla_\nu h + \frac{1}{2}R \nabla_\mu\nabla_\nu h  - \omega \partial_\mu \phi \partial_\nu \phi +\frac{\omega}{2}g_{\mu\nu}\partial_\sigma \phi \partial^\sigma \phi   \nonumber\\
& +\frac{f^2}{4}g_{\mu\nu}F_{\rho\sigma}F^{\rho\sigma} -f^2 F_{\mu\gamma}F_\nu{ }^\gamma=0, \\
\label{c2.scalar1}
& \omega \square \phi =  \frac{\partial_\phi h}{8}G + \frac{f\partial_\phi f}{2}F^{\mu\nu}F_{\mu\nu} ,\\
\label{c2.vector1}
&\frac{ \partial}{\partial x^\mu} \left[\sqrt{-g}f^2 F^{\mu\nu}\right]=0.
\end{align}
Following Refs. \cite{Soda,WFK,Lahiri}, the field configuration for the vector field will be chosen to be $A_\mu =(0,A_x(t),0,0)$, which is compatible with the Bianchi type I metric shown in Eq. (\ref{bianchi-I}). As a result, the following solution of the field equation of vector field (\ref{c2.vector1}) is given by 
\begin{equation}
 \dot A_x = p_A f^{-2} \exp\left[-\alpha -4\sigma \right],
\end{equation}
where $p_A$ is an integration constant \cite{Soda}. Thanks to this solution the  field equations  shown in Eqs. (\ref{c2.Einstein}) and (\ref{c2.scalar1}) now reduce to
\begin{align}
\label{c2.alpha1}
\dot\alpha^2 = ~&\dot\sigma^2 +\frac{\omega \dot\phi^2}{6M_p^2} +\frac{\dot h}{M_p^2} \left(\dot\alpha^3 -2\dot\sigma^3-3\dot\alpha\dot\sigma^2\right) +\frac{f^{-2}}{6 M_p^2} \exp\left[-4\alpha -4\sigma\right]p_A^2, \\
\label{c2.alpha2}
\ddot\alpha =& -3\dot\alpha^2 +\frac{\dot h}{2M_p^2} \left(2\ddot\alpha \dot\alpha -2\ddot\sigma \dot\sigma +5\dot\alpha^3 -9\dot\alpha \dot\sigma^2 -4\dot\sigma^3\right)+\frac{\ddot h}{2M_p^2} \left(\dot\alpha^2-\dot\sigma^2\right) \nonumber\\
&+ \frac{f^{-2}}{6 M_p^2} \exp\left[-4\alpha -4\sigma\right]p_A^2 , \\
\label{c2.sigma}
\ddot\sigma = &-3\dot\alpha \dot\sigma +\frac{ \dot h}{M_p^2} \left[ \ddot\alpha \dot\sigma +\ddot\sigma \left(\dot\alpha+2\dot\sigma\right) + 3 \dot\alpha \dot\sigma \left(\dot\alpha+\dot\sigma\right)  \right] +\frac{\dot\sigma \ddot h }{M_p^2}\left(\dot\alpha+\dot\sigma  \right) \nonumber\\
&+\frac{f^{-2}}{3 M_p^2} \exp\left[-4\alpha -4\sigma\right]p_A^2 ,\\
 \label{c2.phi}
\omega \ddot\phi  = &-3\omega \dot\alpha \dot\phi -3\left(\dot\alpha + \dot\sigma\right) \left[\ddot\alpha \left(\dot\alpha -\dot\sigma\right) -2\ddot\sigma \dot\sigma +\dot\alpha^3 -\dot\alpha\dot\sigma \left(\dot\alpha+2\dot\sigma\right)\right]\partial_\phi h \nonumber\\
&+f^{-3} \partial_\phi f \exp \left[-4\alpha -4\sigma\right]p_A^2 .
\end{align}
These equations are of course consistent with that derived in Ref. \cite{Lahiri}.
Consequently, a set of algebraic equations coming from these field equations is given by
\begin{align}\label{c2.eq-Friedmann}
\zeta^2 &=  \eta^2 +\frac{\omega \xi^2}{6} +2 \left(\zeta^3 -2\eta^3 -3\zeta \eta^2\right)  u +\frac{v}{6}, \\
\label{c2.eq-zeta}
-\zeta &= -3\zeta^2+  \left(5\zeta^3 -4\eta^3- 9\zeta\eta^2-2\zeta^2+2\eta^2 \right) u  + \left(\zeta^2 -\eta^2\right)u+\frac{v}{6} , \\
\label{c2.eq-eta}
-\eta &= -3\zeta \eta + 2  \eta \left(\zeta+\eta\right) \left(3\zeta-1\right)u+\frac{v}{3} ,\\
\label{c2.eq-phi}
-\omega\xi &= -3\omega  \xi\zeta  -3\lambda  \left(\zeta +\eta \right) \left[\zeta^3 -\zeta \left(\zeta-\eta \right) -\zeta\eta \left(\zeta+2\eta\right)+2\eta^2\right] u -\rho v,
\end{align}
here we have kept initial configurations for the scale factors $\alpha$ and $\sigma$ as well as that of scalar field $\phi$ and its function $h(\phi)$ as proposed in the previous section. Additionally, we have introduced an exponential form for the  $f(\phi)$ such as
\begin{equation}
f(\phi) = f_0 \exp \left[- \frac{\rho\phi}{M_p} \right]
\end{equation}
along with an associated variable as
\begin{equation}
v= \frac{p_A^2 f_0^{-2}}{M_p^2} \exp \left[2\rho \phi_0 \right], 
\end{equation}
where $f_0$ and $\rho$ are positive constants. The choice of negative sign in the definition of $f(\phi)$ is necessary for the positive $\rho$ in order to obtain inflationary solutions due to the constraint for $\rho$, 
\begin{equation} \label{c2.constraint1}
-\rho \xi +2\zeta+2\eta =1,
\end{equation}
along with that for $\lambda$,
\begin{equation} \label{c2.constraint2}
\lambda \xi =2.
\end{equation}
It is noted that these constraints appear to make all terms in the field equations proportional to $t^{-2}$. As a result, Eq. (\ref{c2.constraint1}) can be rewritten as
\begin{equation} \label{c2.eq.of.rho.1}
\eta =\frac{1}{2}+\frac{\rho}{\lambda}-\zeta,
\end{equation}
with the help of the constraint equation (\ref{c2.constraint2}). It is clear that if the negative sign in the exponential form of $f(\phi)$ is not present, it will not be easy to obtain large positive values for $\zeta$ provided that $\lambda$ and $\rho$ are positive definite. {\color{black}It becomes clear that $\rho \gg \lambda$ is required to have inflationary solutions with $\zeta \gg 1$}.

As said above, Eq. (\ref{c2.eq.of.rho.1}) is exactly the equation, which Eqs. (\ref{c2.eq-zeta}), (\ref{c2.eq-eta}), and (\ref{c2.eq-phi}) need in order to make a complete set of algebraic equations for $\zeta$, $\eta$, $u$, and $v$. Now, we would like to seek analytical solutions to this set of equations. First, Eqs. (\ref{c2.eq-eta}) and (\ref{c2.eq-phi}) can be solved to give 
\begin{align}\label{c2.eq.of.u}
&u=\frac{2 \lambda  \left[18\lambda  \rho \zeta ^2  - 3 \left(5 \lambda  \rho + 6  \rho ^2 +4 \omega \right)\zeta  +3 \lambda  \rho +6 \rho ^2+4 \omega \right]}{3 (\lambda +2 \rho ) \left[3 \lambda \left(\lambda +6 \rho \right) \zeta ^2 -4 \left( \lambda ^2+5   \lambda  \rho +4   \rho ^2 \right) \zeta+\lambda ^2+6 \lambda  \rho +8 \rho ^2\right]}, \\
\label{c2.eq.of.v}
&v= -\frac{ \left(3 \zeta -1\right) \Omega }{2 \lambda  \left[3 \lambda \left(\lambda +6 \rho \right) \zeta ^2 -4 \left( \lambda ^2+5   \lambda  \rho +4   \rho ^2 \right) \zeta+\lambda ^2+6 \lambda  \rho +8 \rho ^2\right]},
\end{align}
with the help of the solution shown in Eq. (\ref{c2.eq.of.rho.1}). Here, we have defined an additional variable $\Omega$ as
\begin{align}\label{c2.eq.of.omega}
\Omega \equiv ~&18\lambda^2 \left(\lambda+2\rho \right)\zeta^3 -\left(33 \lambda ^3+96 \lambda ^2 \rho +60 \lambda  \rho ^2 {\color{black}-48 \omega \lambda}\right)\zeta^2 \nonumber\\
&+\left(18 \lambda ^3+24 \rho ^3+78 \lambda ^2 \rho +96 \lambda  \rho ^2 {\color{black}-40\omega \lambda -48 \omega\rho}\right)\zeta \nonumber\\
& -3 \lambda ^3-24 \rho ^3-18 \lambda ^2 \rho -36 \lambda  \rho ^2{\color{black}+8\omega \lambda +16\omega \rho}.
\end{align}
As a result, inserting these definitions into either Eq. (\ref{c2.eq-Friedmann}) or Eq. (\ref{c2.eq-zeta}) yields the following non-trivial equation of $\zeta$,
\begin{equation} \label{c2.eq.of.zeta.2}
F(\zeta) \equiv A\zeta^3 +B \zeta^2 +C\zeta +D=0,
\end{equation}
with
\begin{align}
A&=54 \lambda ^4+108 \lambda ^3 \rho,\\
B&=-81 \lambda ^4-252 \lambda ^3 \rho -180 \lambda ^2 \rho ^2{\color{black}+144 \omega \lambda ^2},\\
C&=30 \lambda ^4+156 \lambda ^3 \rho +288 \lambda ^2 \rho ^2{\color{black}-48 \omega \lambda ^2}+192 \lambda  \rho ^3,\\
D&=-3 \lambda ^4 -24 \lambda ^3 \rho -84 \lambda ^2 \rho ^2-144 \lambda  \rho ^3-32 \omega \lambda  \rho-96 \rho ^4-64\omega \rho ^2.
\end{align}
It appears that for the inflationary constraint, $\rho\gg \lambda$,  the coefficients $A$ and $D$ behave approximately as $A\sim 108\lambda^3\rho >0$ and $D \sim -96
\rho^4 <0$. The result that $AD<0$ indicates that Eq. (\ref{c2.eq.of.zeta.2}) will admit at least one positive root $\zeta >0$, which might be used to present inflationary solutions as expected. This conclusion is based on an observation that the curve $F(\zeta)$ will cross the positive horizontal-axis at least one time due to the fact that $F(\zeta) \sim A \zeta^3>0 $ as $\zeta \gg 1$ and $F(\zeta) =D<0$ as $\zeta=0$. And the intersection point is exactly a positive root to the equation $F(\zeta)=0$. 

Furthermore, we are able to estimate the approximated value of the desired solution provided that $\rho \gg \lambda$. Indeed, we can simplify the $F(\zeta)$ by taking leading term of $A$, $B$, $C$, and $D$ as follows
\begin{align}
F(\zeta) \simeq \tilde F(\zeta)& \equiv  108 \lambda ^3 \rho \zeta^3 -180 \lambda ^2 \rho ^2\zeta^2 + 192 \lambda  \rho ^3 \zeta -96 \rho ^4 \nonumber\\
&= 12 \lambda^3 \rho \left( 9 \zeta^3 - 15 \frac{ \rho }{\lambda} \zeta^2 + 16  \frac{ \rho ^2}{\lambda^2} \zeta - 8 \frac{\rho ^3}{\lambda^3}\right).
\end{align}
As a result, the equation $\tilde F(\zeta)=0$ can be solved to give a non-trivial solution, which does not depend on the property (canonical or phantom) of scalar field, as
\begin{equation}
 \zeta _0 = \frac{1}{9}\frac{\rho}{\lambda} \left(5 +\sqrt[3]{18 \sqrt{62} +89}- \frac{23 }{\sqrt[3]{18 \sqrt{62}+89}}\right) \simeq 0.82\frac{\rho}{\lambda} \gg 1.
\end{equation}
{\color{black}Note that we are also  able to show by exact method(s) that the cubic polynomial in Eq. (\ref{c2.eq.of.zeta.2}) has only one real root which can be approximated to be $0.82 \rho/ \lambda$. However, we will not present the proof here due to its lengthy calculation since the coefficients $A$, $B$, $C$, and $D$ of Eq. (\ref{c2.eq.of.zeta.2}) are quite complicated.
Instead, we will show that $\zeta_0$ is indeed the only real solution to the equation, $\tilde F(\zeta)=0$, during the inflationary phase with $\rho/\lambda \gg 1$ using the well-known Cardano's or Vieta's method. Indeed, it is straightforward to define the following discriminant
\begin{equation}
\Delta = - 26784 \frac{\rho^6}{\lambda^6}.
\end{equation}
It is known that if $\Delta <0$ then the cubic equation, $\tilde F(\zeta)=0$, admits one real root and two complex roots. And it is straightforward to see that $\Delta$ is always negative definite. This proves that $\zeta_0$ is the only real solution to the cubic equation, $\tilde F(\zeta)=0$, as expected.}

To be more specific, let us make a simple comparison between the exact solution and the approximated one with a particular value of $\lambda$ and $\rho$. For example, it is straightforward to have $\zeta_0 \simeq 41.02 $ for $\lambda =1$ and $\rho=50$, while the exact solution of $F(\zeta) =0$ turns out to be $\zeta \simeq 41.44$. It is clear that the gap between $\zeta_0$ and $\zeta$ is tiny compared to their actual value. Hence, the approximated solution of $\zeta=\zeta_0$ is acceptable so that we can estimate, according to Eq. (\ref{c2.eq.of.rho.1}), the corresponding $\eta$ as follows
\begin{equation}
\eta =\frac{1}{2}+\frac{\rho}{\lambda}-\zeta \simeq 0.18 \frac{\rho}{\lambda}.
\end{equation}
It is clear that the corresponding  anisotropy parameter is given by
\begin{equation}
\Sigma \equiv \frac{\eta}{\zeta} \simeq 0.22.
\end{equation}
This value of $\Sigma$ is of the same order as that obtained in the previous section without the presence of the coupling $f^2(\phi)F^2/4$, for instance see Eq. (\ref{ratio}) or Fig. \ref{fig3}, in contrast to our expectation that its magnitude would be reduced to a small number due to the existence of vector field.  The reason might be due to the absence of the potential $V(\phi)$. Indeed, Ref. \cite{Lahiri} has investigated a different scenario, in which $V(\phi)$ shows up, and shown that anisotropic inflationary solutions with small hairs can exist. 
\section{CONCLUSIONS} \label{sec4}
We have shown that the scalar-Gauss-Bonnet model without the pure potential of scalar field $V(\phi)$ can admit the Bianchi type I metric as its anisotropic inflationary solution if the scalar field $\phi$ acts as the phantom field with $\omega =-1$. However, the obtained spatial anisotropy turns out to be large, in contrast to our expectation and the other models such as the KSW model \cite{Soda,Soda1,WFK}. Hence, we have introduced the coupling between the scalar and vector fields such as $f^2(\phi)F_{\mu\nu}F^{\mu\nu}$ with the hope that the large spatial hair would be reduced to small one, following the investigations in Refs. \cite{Soda,Soda1,WFK} as well as Ref. \cite{Lahiri}.  It turns out, however, that the extended value of the spatial hair is still large, regardless of the property of scalar field. This result is indeed in contrast to that obtained in Ref. \cite{Lahiri}, where $V(\phi)$ is taken into account.  The results obtained in the present paper indicate that the scalar-Gauss-Bonnet gravity model with the absence of potential of scalar field might not be suitable to produce a small anisotropic hair during the inflationary phase. In other words, the present study indicates that the existence of $V(\phi)$ might not be trivially ignored in the context of anisotropic Gauss-Bonnet  inflation. This point is in agreement with the recent investigation in Ref. \cite{no-go} that the scalar-Gauss-Bonnet model without the inflaton potential might not be  viable. We hope that our research would shed more light on the cosmological implications of the Gauss-Bonnet term.
\section*{APPENDIX} 
In this Appendix, we will list the explicit expressions of non-vanishing components of the Riemann tensor, $R^{\alpha}{ }_{\nu\rho\sigma}$, and the Ricci tensor, $R_{\mu\nu}\equiv R^\rho{ }_{\mu\rho\nu}$, for the Bianchi type I metric given by Eq. (\ref{bianchi-I}). Firstly, we will list  the following non-vanishing Christoffel symbols,
\begin{equation}
\Gamma^0_{11}=g_{11}\left(\dot\alpha-2\dot\sigma\right), ~\Gamma^0_{22}=\Gamma^0_{33}=g_{33}\left(\dot\alpha+\dot\sigma\right),~\Gamma^1_{10}=\dot\alpha-2\dot\sigma, ~ \Gamma^2_{20}=\Gamma^3_{30}=\dot\alpha+\dot\sigma.
\end{equation}
As a result, the following non-zero components of Riemann tensor, $R^{\mu}{ }_{\nu\rho\sigma}$ can be defined to be
\begin{align}
R^0{ }_{101}&=g_{11}\left[\ddot\alpha -2\ddot\sigma+\left(\dot\alpha -2\dot\sigma\right)^2\right], \nonumber\\
R^0{ }_{202}&=R^0{ }_{303}=  g_{33} \left[\ddot\alpha +\ddot\sigma+\left(\dot\alpha +\dot\sigma\right)^2\right], \nonumber\\
R^{1}{ }_{010}&= -\ddot\alpha +2\ddot\sigma -\left(\dot\alpha-2\dot\sigma\right)^2,\nonumber\\
R^2{ }_{020}&=R^3{ }_{030}=-\ddot\alpha -\ddot\sigma - \left(\dot\alpha+\dot\sigma\right)^2, \nonumber\\
R^1{ }_{212}&=R^1{ }_{313}=g_{33} \left(\dot\alpha -2\dot\sigma\right) \left(\dot\alpha +\dot\sigma\right), \nonumber\\
R^2{ }_{121}&=R^3{ }_{131}= g_{11} \left(\dot\alpha -2\dot\sigma\right) \left(\dot\alpha +\dot\sigma\right), \nonumber\\
R^2{ }_{323}&=R^3{ }_{232} = g_{33} \left(\dot\alpha +\dot\sigma\right)^2.
\end{align}
Hence, it is straightforward to obtain the following Ricci tensor, $R_{\mu\nu}\equiv R^\rho{ }_{\mu\rho\nu}$ as follows
\begin{align}
R_{00}&=-3 \left(\ddot\alpha +\dot\alpha^2 +2\dot\sigma^2 \right), \nonumber\\
R_{11}&= g_{11}\left(\ddot\alpha -2\ddot\sigma +3\dot\alpha^2 -6\dot\alpha \dot\sigma \right), \nonumber\\
R_{22}&=R_{33}=g_{33}\left(\ddot\alpha +\ddot\sigma +3\dot\alpha^2 +3\dot\alpha\dot\sigma\right).
\end{align}
Finally, the corresponding Ricci scalar turns out to be
\begin{equation}
R \equiv g^{\mu\nu}R_{\mu\nu} =6 \left(\ddot\alpha +2\dot\alpha^2 +\dot\sigma^2 \right)
\end{equation}
along with the non-trivial components of the Einstein tensor, $G_{\mu\nu}= R_{\mu\nu}-\frac{1}{2}g_{\mu\nu}R$,  given by
\begin{align}
G_{00}&= 3 \left(\dot\alpha^2 -\dot\sigma^2 \right), \nonumber\\
G_{11}&=-g_{11} \left[2\left(\ddot\alpha +\ddot\sigma \right)+3 \left(\dot\alpha+\dot\sigma\right)^2\right], \nonumber\\
G_{22}&=G_{33}=-g_{33} \left[2\ddot\alpha -\ddot\sigma +3\left(\dot\alpha^2-\dot\alpha \dot\sigma+\dot\sigma^2\right)\right].
\end{align}
Given these results, we will be able to define the Gauss-Bonnet term to be
\begin{equation} \label{RGB}
G = 24 \left(\dot\alpha + \dot\sigma\right) \left[\ddot\alpha \left(\dot\alpha -\dot\sigma\right) -2\ddot\sigma \dot\sigma +\dot\alpha^3 -\dot\alpha\dot\sigma \left(\dot\alpha+2\dot\sigma\right)\right].
\end{equation}
It is clear that in the isotropic limit, $\sigma \to 0$, then all above definitions will become to that shown in Ref. ~\cite{darkenergy}. 

As a result, the following non-vanishing {00}, {11}, and {22 (33)} components of the Einstein field equations (\ref{c1.Einstein}) can be defined to be
\begin{align} \label{00}
3M_p^2 \left(\dot\alpha^2 -\dot\sigma^2\right) =&-3 \dot h \left(\dot\alpha^3 -2\dot\sigma^3-3\dot\alpha\dot\sigma^2\right) +\frac{\dot\phi^2}{2}, \\
\label{11}
M_p^2 \left[2\left(\ddot\alpha +\ddot\sigma \right)+3 \left(\dot\alpha+\dot\sigma\right)^2\right] =& -\frac{\dot\phi^2}{2}-\ddot h \left(\dot\alpha+\dot\sigma\right)^2 \nonumber\\
&-\dot h \left(3\ddot\alpha \dot\alpha +6\ddot\sigma \dot\sigma +5\dot\alpha^3+2\dot\sigma^3 -6\dot\alpha^2 \dot\sigma+18\dot\alpha \dot\sigma^2\right) ,\\
\label{33}
M_p^2 \left[2\ddot\alpha -\ddot\sigma +3\left(\dot\alpha^2-\dot\alpha \dot\sigma+\dot\sigma^2\right)\right] =&-\frac{\dot\phi^2}{2}-\ddot h \left(\dot\alpha^2 -\dot\alpha \dot\sigma -2\dot\sigma^2\right) \nonumber\\
& - \dot h \left(3\ddot\alpha \dot\alpha  -3\ddot\sigma \dot\sigma +5\dot\alpha^3+2\dot\sigma^3 +3\dot\alpha^2 \dot\sigma   \right) ,
\end{align}
respectively. It is clear that $00$-component equation (\ref{00}) is identical to Eq. (\ref{c1.alpha1}), which is called the Friedmann equation. 
As a result, eliminating $\ddot\alpha$ in both Eqs. (\ref{11}) and (\ref{33}) leads to the anisotropy equation  (\ref{c1.sigma}). On the other hand, eliminating $\ddot\sigma$ in both Eqs. (\ref{11}) and (\ref{33})  leads to Eq. (\ref{c1.alpha2}) with the help of the Friedmann equation (\ref{00}).
\section*{ACKNOWLEDGMENT}
This research is supported by the Vietnam National Foundation for Science and Technology Development (NAFOSTED) under Grant No. 103.01-2017.12. We would like to thank an anonymous referee very much for useful comments. T.Q.D. is deeply grateful to Professor W. F. Kao of Institute of Physics in National Chiao Tung University for his useful advice on anisotropic inflation. T.Q.D. would like to thank his colleagues, Dr. N. T. T. Nhan, Dr. N. T. Cuong, and Mr. N. C. Viet,  very much for their useful help. 



\begin{thebibliography}{99} 
\bibitem{Saadeh:2016sak} 
  D.~Saadeh, S.~M.~Feeney, A.~Pontzen, H.~V.~Peiris, and J.~D.~McEwen,
 {\it Phys.\ Rev.\ Lett.\ }{\bf 117} (2016)  131302 
  [arXiv:1605.07178];
J. Soltis, A. Farahi, D. Huterer, and C. M. Liberato II,
 {\it Phys. Rev. Lett.\ }{\bf 122} (2019) 091301 
 [arXiv:1902.07189].

\bibitem{GH} 
  G.~W.~Gibbons and S.~W.~Hawking,
   {\it Phys.\ Rev.\ D} {\bf 15} (1977) 2738;
  S.~W.~Hawking and I.~G.~Moss,
  {\it Phys.\ Lett.\ B} {\bf 110} (1982) 35.

\bibitem{wald} 
  R.~M.~Wald,
 {\it Phys.\ Rev.\ D} {\bf 28} (1983) 2118.

\bibitem{Barrow:1987ia} 
  J.~D.~Barrow,
  {\it Phys.\ Lett.\ B} {\bf 187} (1987) 12;
  Y.~Kitada and K.~i.~Maeda,
  {\it Phys.\ Rev.\ D} {\bf 45} (1992) 1416.

\bibitem{inhomogeneous}
  M.~Kleban and L.~Senatore,
  {\it J. Cosmol. Astropart. Phys.} {\bf 10} (2016) 022
 [arXiv:1602.03520]; 
  W.~E.~East, M.~Kleban, A.~Linde, and L.~Senatore,
  {\it J. Cosmol. Astropart. Phys.} {\bf 09} (2016) 010
[arXiv:1511.05143].

\bibitem{Carroll:2017kjo} 
  S.~M.~Carroll and A.~Chatwin-Davies,
{\it Phys. Rev. D} {\bf 97} (2018) 046012  [arXiv:1703.09241].

\bibitem{guth}
  A.~H.~Guth,
 {\it Phys.\ Rev.\ D} {\bf 23} (1981) 347;
  A.~D.~Linde,
 {\it Phys.\ Lett.\ B} {\bf 108} (1982) 389;
  A.~A.~Starobinsky,
  {\it Phys.\ Lett.\ B} {\bf 91} (1980) 99;
  A.~Albrecht and P.~J.~Steinhardt,
{\it  Phys.\ Rev.\ Lett.\ }{\bf 48} (1982) 1220;
  A.~D.~Linde,
 {\it Phys.\ Lett.\ B} {\bf 129} (1983) 177.

\bibitem{WMAP} 
  E.~Komatsu {\it et al.} [WMAP Collaboration],
  {\it Astrophys.\ J.\ Suppl.\ }{\bf 192} (2011) 18 
  [arXiv:1001.4538];
  G.~Hinshaw {\it et al.} [WMAP Collaboration],
{\it Astrophys.\ J.\ Suppl.\ }{\bf 208} (2013) 19 
  [arXiv:1212.5226].

\bibitem{Planck}
  P.~A.~R.~Ade {\it et al.} [Planck Collaboration],
{\it Astron.\ Astrophys.\ }{\bf 571} (2014) A22 
  [arXiv:1303.5082];
 {\it Astron.\ Astrophys.\ }{\bf 571} (2014) A23 
  [arXiv:1303.5083].

\bibitem{Bianchi}
  G.~F.~R.~Ellis and M.~A.~H.~MacCallum,
 {\it Commun.\ Math.\ Phys.\ }{\bf 12} (1969) 108;
  G.~F.~R.~Ellis,
  {\it Gen.\ Rel.\ Grav.\ }{\bf 38} (2006) 1003.

\bibitem{Pitrou:2008gk} 
  C.~Pitrou, T.~S.~Pereira, and J.~P.~Uzan,
 {\it J. Cosmol. Astropart. Phys.} {\bf 04} (2008) 004 
  [arXiv:0801.3596];
  A.~E.~Gumrukcuoglu, C.~R.~Contaldi, and M.~Peloso,
{\it J. Cosmol. Astropart. Phys.} {\bf 07} (2007) 005 
  [arXiv:0707.4179].

\bibitem{Buchert:2015wwr} 
  T.~Buchert, A.~A.~Coley, H.~Kleinert, B.~F.~Roukema, and D.~L.~Wiltshire,
{\it  Int.\ J.\ Mod.\ Phys.\ D} {\bf 25} (2016) 1630007 
  [arXiv:1512.03313].

\bibitem{Soda} 
  S.~Kanno, J.~Soda, and M.~a.~Watanabe,
{\it J. Cosmol. Astropart. Phys.}  {\bf 12} (2010)  024
  [arXiv:1010.5307];
M.~a.~Watanabe, S.~Kanno, and J.~Soda,
 {\it Phys.\ Rev.\ Lett.\ }{\bf 102} (2009) 191302 
  [arXiv:0902.2833].

\bibitem{Soda1}
  A.~Maleknejad, M.~M.~Sheikh-Jabbari, and J.~Soda,
 {\it Phys.\ Rept.\ }{\bf 528} (2013) 161 
  [arXiv:1212.2921];
  J.~Soda,
 {\it  Class.\ Quant.\ Grav.\ }{\bf 29} (2012) 083001 
  [arXiv:1201.6434].

\bibitem{WFK}
  T.~Q.~Do and W.~F.~Kao,
{\it Phys.\ Rev.\ D} {\bf 84} (2011) 123009;
  T.~Q.~Do and W.~F.~Kao,
 {\it  Class.\ Quant.\ Grav.\ }{\bf 33} (2016) 085009;
  T.~Q.~Do and W.~F.~Kao,
 {\it Phys.\ Rev.\ D} {\bf 96} (2017) 023529.

\bibitem{WFK1}
  T.~Q.~Do, W.~F.~Kao, and I.~C.~Lin,
 {\it Phys.\ Rev.\ D} {\bf 83} (2011) 123002;
  T.~Q.~Do and S.~H.~Q.~Nguyen,
{\it  Int.\ J.\ Mod.\ Phys.\ D} {\bf 26} (2017) 1750072
  [arXiv:1702.08308].


\bibitem{extensions}
  R.~Emami, H.~Firouzjahi, S.~M.~Sadegh Movahed, and M.~Zarei,
  {\it J. Cosmol. Astropart. Phys.} {\bf 02} (2011)  005
  [arXiv:1010.5495];
  K.~Murata and J.~Soda,
  {\it J. Cosmol. Astropart. Phys.} {\bf 06} (2011) 037
  [arXiv:1103.6164];
  S.~Hervik, D.~F.~Mota, and M.~Thorsrud,
  {\it J. High Energy Phys.} {\bf 11} (2011) 146
  [arXiv:1109.3456];
  K.~Yamamoto, M.~a.~Watanabe, and J.~Soda,
 {\it Class.\ Quantum\ Grav.} {\bf 29} (2012) 145008
  [arXiv:1201.5309];
  M.~Thorsrud, D.~F.~Mota, and S.~Hervik,
  {\it J. High Energy Phys.} {\bf 10}  (2012) 066
  [arXiv:1205.6261];
  A.~Maleknejad and M.~M.~Sheikh-Jabbari,
 {\it Phys.\ Rev.\ D} {\bf 85} (2012) 123508 
  [arXiv:1203.0219];
  K.~i.~Maeda and K.~Yamamoto,
  {\it Phys.\ Rev.\ D} {\bf 87} (2013) 023528
  [arXiv:1210.4054];
  J.~Ohashi, J.~Soda, and S.~Tsujikawa,
 {\it  Phys.\ Rev.\ D} {\bf 87} (2013)  083520
  [arXiv:1303.7340];
  J.~Ohashi, J.~Soda, and S.~Tsujikawa,
 {\it Phys.\ Rev.\ D} {\bf 88} (2013) 103517
  [arXiv:1310.3053];
  A.~Ito and J.~Soda,
  {\it Phys.\ Rev.\ D} {\bf 92} (2015)  123533 
  [arXiv:1506.02450];
  A.~A.~Abolhasani, M.~Akhshik, R.~Emami, and H.~Firouzjahi,
  {\it J. Cosmol. Astropart. Phys.} {\bf 03} (2016) 020
  [arXiv:1511.03218];
  M.~Karciauskas,
  {\it Mod.\ Phys.\ Lett.\ A} {\bf 31} (2016)  1640002
  [arXiv:1604.00269];
  M.~Tirandari and K.~Saaidi,
 {\it Nucl.\ Phys.\ B} {\bf 925} (2017) 403 
  [arXiv:1701.06890];
  A.~Ito and J.~Soda,
{\it  Eur.\ Phys.\ J.\ C} {\bf 78} (2018)  55 
  [arXiv:1710.09701];
  T.~Fujita and I.~Obata,
{\it J. Cosmol. Astropart. Phys.} {\bf 01} (2018)  049 
  [arXiv:1711.11539];
  J.~Holland, S.~Kanno, and I.~Zavala,
 {\it Phys.\ Rev.\ D} {\bf 97} (2018) 103534 
  [arXiv:1711.07450];
  T.~Q.~Do and W.~F.~Kao,
{\it  Eur.\ Phys.\ J.\ C} {\bf 78} (2018) 360 
  [arXiv:1712.03755];
  T.~Q.~Do and W.~F.~Kao,
  {\it Eur.\ Phys.\ J.\ C} {\bf 78} (2018) 531.

\bibitem{primary}
  B.~Zwiebach,
 {\it Phys.\ Lett.\ B} {\bf 156} (1985) 315;
  D.~G.~Boulware and S.~Deser,
 {\it Phys.\ Rev.\ Lett.\ }{\bf 55} (1985) 2656;
  D.~J.~Gross and E.~Witten,
{\it  Nucl.\ Phys.\ B} {\bf 277} (1986) 1;
  D.~J.~Gross and J.~H.~Sloan,
 {\it  Nucl.\ Phys.\ B} {\bf 291} (1987) 41;
  R.~R.~Metsaev and A.~A.~Tseytlin,
 {\it Nucl.\ Phys.\ B} {\bf 293} (1987) 385;
  R.~R.~Metsaev and A.~A.~Tseytlin,
{\it Phys.\ Lett.\ B} {\bf 191} (1987) 354.

\bibitem{Lovelock:1971yv} 
  D.~Lovelock,
 {\it J.\ Math.\ Phys.\ }{\bf 12} (1971) 498.

\bibitem{ART}
  I.~Antoniadis, J.~Rizos, and K.~Tamvakis,
 {\it Nucl.\ Phys.\ B} {\bf 415} (1994) 497 
  [hep-th/9305025]; 
  J.~Rizos and K.~Tamvakis,
 {\it Phys.\ Lett.\ B} {\bf 326} (1994) 57 
  [gr-qc/9401023];
  P.~Kanti, J.~Rizos, and K.~Tamvakis,
 {\it Phys.\ Rev.\ D} {\bf 59} (1999) 083512 
  [gr-qc/9806085].

\bibitem{KSS}
  S.~Kawai, M.~a.~Sakagami, and J.~Soda,
 {\it Phys.\ Lett.\ B} {\bf 437}  (1998) 284
  [gr-qc/9802033];
  S.~Kawai and J.~Soda,
 {\it Phys.\ Lett.\ B} {\bf 460} (1999) 41 
  [gr-qc/9903017].

\bibitem{bianchi-I}
  A.~Toporensky and S.~Tsujikawa,
{\it  Phys.\ Rev.\ D} {\bf 65} (2002) 123509 
  [gr-qc/0202067].

\bibitem{darkenergy}
  S.~Nojiri, S.~D.~Odintsov, and M.~Sasaki,
 {\it Phys.\ Rev.\ D} {\bf 71}  (2005) 123509
  [hep-th/0504052].


\bibitem{review-darkenergy}
  S.~Nojiri and S.~D.~Odintsov,
{\it  Phys.\ Lett.\ B} {\bf 631} (2005) 1 
  [hep-th/0508049];
  S.~Nojiri, S.~D.~Odintsov, and O.~G.~Gorbunova,
{\it  J.\ Phys.\ A} {\bf 39} (2006) 6627 
  [hep-th/0510183];
  B.~M.~N.~Carter and I.~P.~Neupane,
  {\it J. Cosmol. Astropart. Phys.} { 06} (2006) 004 
  [hep-th/0512262];
  G.~Cognola, E.~Elizalde, S.~Nojiri, S.~D.~Odintsov, and S.~Zerbini,
{\it  Phys.\ Rev.\ D} {\bf 73} (2006) 084007 
  [hep-th/0601008];
  S.~Nojiri and S.~D.~Odintsov,
{\it  Int.\ J.\ Geom.\ Meth.\ Mod.\ Phys.\ }  {\bf 4}  (2007) 115
  [hep-th/0601213];
  T.~Koivisto and D.~F.~Mota,
  {\it Phys.\ Lett.\ B} {\bf 644} (2007) 104 
  [astro-ph/0606078];
  G.~Cognola, E.~Elizalde, S.~Nojiri, S.~Odintsov, and S.~Zerbini,
{\it  Phys.\ Rev.\ D} {\bf 75} (2007) 086002 
  [hep-th/0611198];
  B.~Li, J.~D.~Barrow, and D.~F.~Mota,
 {\it Phys.\ Rev.\ D} {\bf 76} (2007) 044027 
  [arXiv:0705.3795].

\bibitem{inflation} 
  P.~Kanti, R.~Gannouji, and N.~Dadhich,
 {\it Phys.\ Rev.\ D} {\bf 92} (2015) 041302(R)  [arXiv:1503.01579];
  P.~Kanti, R.~Gannouji, and N.~Dadhich,
 {\it Phys.\ Rev.\ D} {\bf 92} (2015)  083524 
  [arXiv:1506.04667];
  O.~P.~Santillan,
{\it J. Cosmol. Astropart. Phys.} {\bf 07} (2017) 008
  [arXiv:1703.01713];
  T.~Anson, E.~Babichev, C.~Charmousis, and S.~Ramazanov,
  arXiv:1903.02399.

\bibitem{observational-inflation}
  Z.~K.~Guo, N.~Ohta, and S.~Tsujikawa,
{\it  Phys.\ Rev.\ D} {\bf 75}  (2007) 023520
  [hep-th/0610336];
  Z.~K.~Guo and D.~J.~Schwarz,
 {\it  Phys.\ Rev.\ D} {\bf 80} (2009) 063523 
  [arXiv:0907.0427]; 
  Z.~K.~Guo and D.~J.~Schwarz,
{\it  Phys.\ Rev.\ D} {\bf 81} (2010) 123520 
  [arXiv:1001.1897]; 
  P.~X.~Jiang, J.~W.~Hu, and Z.~K.~Guo,
{\it  Phys.\ Rev.\ D} {\bf 88} (2013) 123508 
  [arXiv:1310.5579];
  S.~Koh, B.~H.~Lee, W.~Lee, and G.~Tumurtushaa,
{\it  Phys.\ Rev.\ D} {\bf 90} (2014)  063527 
  [arXiv:1404.6096];
  C.~van de Bruck, K.~Dimopoulos, and C.~Longden,
{\it  Phys.\ Rev.\ D} {\bf 94} (2016) 023506 
  [arXiv:1605.06350];
  S.~Chakraborty, T.~Paul, and S.~SenGupta,
 {\it Phys.\ Rev.\ D} {\bf 98} (2018) 083539 
  [arXiv:1804.03004];
  S.~D.~Odintsov and V.~K.~Oikonomou,
 {\it Phys.\ Rev.\ D} {\bf 98} (2018) 044039 
  [arXiv:1808.05045].

\bibitem{Antoniou:2017acq} 
  G.~Antoniou, A.~Bakopoulos, and P.~Kanti,
 {\it Phys.\ Rev.\ Lett.\ }{\bf 120}  (2018) 131102
  [arXiv:1711.03390];
  D.~D.~Doneva and S.~S.~Yazadjiev,
 {\it Phys.\ Rev.\ Lett.\ }{\bf 120} (2018) 131103 
  [arXiv:1711.01187];
  H.~O.~Silva, J.~Sakstein, L.~Gualtieri, T.~P.~Sotiriou, and E.~Berti,
 {\it Phys.\ Rev.\ Lett.\ }{\bf 120} (2018) 131104 
  [arXiv:1711.02080].


\bibitem{blackhole}
  P.~Kanti, N.~E.~Mavromatos, J.~Rizos, K.~Tamvakis, and E.~Winstanley,
 {\it Phys.\ Rev.\ D} {\bf 54} (1996) 5049 
  [hep-th/9511071];
  P.~Kanti, N.~E.~Mavromatos, J.~Rizos, K.~Tamvakis, and E.~Winstanley,
{\it  Phys.\ Rev.\ D} {\bf 57} (1998) 6255 
  [hep-th/9703192];
  T.~Torii, H.~Yajima, and K.~i.~Maeda,
 {\it Phys.\ Rev.\ D} {\bf 55} (1997) 739 
  [gr-qc/9606034];
  K.~D.~Kokkotas, R.~A.~Konoplya, and A.~Zhidenko,
{\it Phys. Rev. D} {\bf 96}  (2017) 064004
 [arXiv:1706.07460].

\bibitem{higherdim-blackhole}
  R.~A.~Konoplya and A.~Zhidenko,
{\it  Phys.\ Rev.\ D} {\bf 95} (2017) 104005 
  [arXiv:1701.01652]; 
  N.~Deppe, A.~Kolly, A.~Frey, and G.~Kunstatter,
 {\it Phys.\ Rev.\ Lett.\ } {\bf 114} (2015) 071102 
  [arXiv:1410.1869];
  R.~A.~Konoplya and A.~Zhidenko,
{\it  Phys.\ Rev.\ D} {\bf 77} (2008) 104004 
  [arXiv:0802.0267];
  R.~G.~Cai,
 {\it Phys.\ Rev.\ D} {\bf 65} (2002) 084014 
  [hep-th/0109133];
  M.~Cvetic, S.~Nojiri, and S.~D.~Odintsov,
{\it  Nucl.\ Phys.\ B} {\bf 628} (2002) 295 
  [hep-th/0112045].

\bibitem{wormhole}
  P.~Kanti, B.~Kleihaus, and J.~Kunz,
{\it  Phys.\ Rev.\ Lett.\ } {\bf 107} (2011) 271101 
  [arXiv:1108.3003];
  P.~Kanti, B.~Kleihaus, and J.~Kunz,
{\it  Phys.\ Rev.\ D} {\bf 85} (2012) 044007 
  [arXiv:1111.4049].

\bibitem{Lahiri} 
  S.~Lahiri,
{\it J. Cosmol. Astropart. Phys.} {\bf 09}  (2016) 025
  [arXiv:1605.09247].

\bibitem{total-derivative}
  T.~P.~Sotiriou and S.~Y.~Zhou,
{\it Phys.\ Rev.\ D} {\bf 90} (2014) 124063 
  [arXiv:1408.1698];
  S.~Nojiri, S.~D.~Odintsov, and V.~K.~Oikonomou,
  {\it Phys.\ Rev.\ D} {\bf 99} (2019) 044050
  [arXiv:1811.07790].

\bibitem{Naruko:2014bxa} 
  A.~Naruko, E.~Komatsu, and M.~Yamaguchi,
  {\it J. Cosmol. Astropart. Phys.} {\bf 04} (2015) 045
  [arXiv:1411.5489].

\bibitem{no-go} 
  G.~Hikmawan, J.~Soda, A.~Suroso, and F.~P.~Zen,
{\it  Phys.\ Rev.\ D} {\bf 93} (2016) 068301 
  [arXiv:1512.00222].
\end{thebibliography}

\end{document}